\documentclass[twocolumn,secnumarabic,amssymb, nobibnotes, aps, prd,superscriptaddress]{revtex4-2}
\usepackage{graphicx}
\usepackage{color}
\usepackage{amsmath}
\usepackage{gensymb}
\usepackage{tabularx}
\usepackage{multirow}

\usepackage[left,modulo]{lineno}
\usepackage{upgreek}

\begin{document}
\onecolumngrid
This article may be downloaded for personal use only. Any other use requires prior permission of the author and AIP Publishing. This article appeared in Marceau Hénot; Computing dielectric spectra in molecular dynamics simulations: Using a cavity to disentangle self and cross correlations. J. Chem. Phys. 28 September 2025; 163 (12): 124501. and may be found at \url{https://doi.org/10.1063/5.0289314}.

\title{Computing dielectric spectra in molecular dynamics simulations: using a cavity to disentangle self and cross correlations}

\author{Marceau H\'enot}
\email[Corresponding author: ]{marceau.henot@cea.fr}
\affiliation{SPEC, CEA, CNRS, Université Paris-Saclay, CEA Saclay Bat 772, 91191 Gif-sur-Yvette Cedex, France.}

\begin{abstract}
Dielectric spectra are typically obtained in molecular dynamics (MD) simulations by analyzing the fluctuations, in the absence of an applied electric field, of the total dipole moment of the simulation box. We compare this standard method with a protocol that focuses on a virtual cavity whose size is chosen to include short-range dipolar cross-correlations, while excluding long-range correlations that are affected by the choice of electrostatic boundary conditions. We tested this protocol on three non-polarizable systems with different dielectric permittivities. We showed that it produces the same dielectric spectra as the standard method while being less sensitive to noise. The question of the decomposition of a dielectric spectrum into self and cross contributions is discussed in the context of both methods. We propose that, for a liquid with a sufficiently high dielectric permittivity, the cavity protocol yields a self-spectrum consistent with the electrostatic boundary conditions applicable to the experimental situation.

\end{abstract}

\maketitle

\section{Introduction}
Dielectric spectroscopy is a popular technique for investigating molecular relaxation mechanisms in polar liquids~\cite{DavidsonCole1951, kremer2002broadband}. At the molecular scale, it probes the response to an electric field of a microscopic volume containing a significant number of molecules~\cite{bottcher1973} within which dipole orientations $\vec{\mu}_i$ may be correlated. In the absence of electric field, these correlations are characterized by the Kirkwood factor~\cite{kirkwood1939dielectric}, defined in its time-dependent generalized form~\cite{Brot1975, bottcher1978}, by:
\begin{align}
       g_\mathrm{K}(t) = \frac{1}{\mu^2}\left\langle \vec{\mu}_i(0) \cdot \sum_{j} \vec{\mu}_j(t) \right\rangle
       \label{eq_def_gKt}
\end{align}
where $\langle.\rangle$ is an ensemble average. Linking the molecular dynamics at the microscopic scale to the frequency-dependent dielectric permittivity $\epsilon(\omega)$, a macroscopic quantity, is not straightforward. This is due to the distinction between the applied electric field and the one felt by the microscopic volume. Taking this into account leads to~\cite{fatuzzo1967theory, Rival1969, bottcher1978}:
\begin{align}
    \frac{(\epsilon(\omega)-\epsilon_\infty)(2\epsilon(\omega)+\epsilon_\infty)}{\epsilon(\omega)(\epsilon_\infty+2)^2} = \frac{\lambda }{3}(g_\mathrm{K}-i\omega\mathcal{L}_{i\omega}[g_\mathrm{K}(t)])
    \label{eq_epsw_gKt}
\end{align}
where $g_\mathrm{K}=g_\mathrm{K}(t=0)$ is the static Kirkwood factor, $\epsilon_\infty$ is the high frequency dielectric constant,  $\mathcal{L}_{i\omega}$ is the one-sided Fourier transform, and $\lambda=\mu^2/(3\epsilon_0k_\mathrm{B}Tv)$ is a dimensionless parameter characterizing the strength of dipole-dipole interaction, with $v$ the molecular volume, $T$ the temperature, $k_\mathrm{B}$ the Boltzmann constant, $\epsilon_0$ the vacuum permittivity.  

In polar liquids, static correlations can significantly contribute to the dielectric permittivity $\epsilon=\epsilon(0)$. For quite a number of liquids, dipolar cross-correlations are positive ($g_\mathrm{K}>1$)~\cite{bottcher1973}, which is generally associated with a tendency for close dipoles to align, although polarizability effects can in some cases make this picture too simple~\cite{dejardin2022kinetic, dejardin2022temperature}. Recently, theoretical development~\cite{dejardin2019linear}, as well as the combination of different experimental techniques~\cite{weigl2019local, gabriel2017debye, koperwas2024experimental} probing molecular reorientation, led to the realization that the dynamics of the correlated volume of dipoles probed by dielectric spectroscopy could significantly differ from the one of individual molecules~\cite{pabst2021generic} and have a large effect~\cite{gabriel2020intermolecular, bohmer2024spectral, bohmer2024dipolar}, or even dominate~\cite{gabriel2017debye, gabriel2018nature}, dielectric spectra. This raises the question of the importance of dipolar correlations in the structural relaxation of molecular glasses \cite{moch2022nongeneric, arrese2025correlation}.
\begin{figure}[h]
\includegraphics[width=\linewidth]{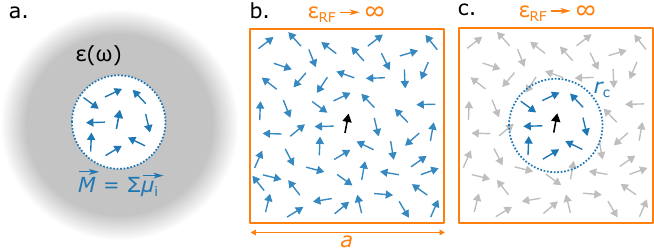}
\caption{\label{fig:fig0}(a) Modeling of the experimental problem: a microscopic volume of total dipole moment $\vec{M}(t)$ is embedded in an infinite continuous medium of permittivity $\epsilon(\omega)$. (b) Configuration of the simulation: the box is embedded into a conducting medium $\epsilon_\mathrm{RF}\rightarrow\infty$. (c) Small virtual cavity of radius $r_\mathrm{c}$ in the simulation box.}
\end{figure}

Molecular dynamics (MD) simulations are a useful tool to complement experimental investigations regarding the molecular dynamics of liquids~\cite{saiz2000dielectric, guiselin2022microscopic, pabst2025glassy}. More specifically, the question of dipolar correlations seems well suited as the sum of eq.~\ref{eq_def_gKt} can be split to study in detail the contribution of self and cross-correlations~\cite{affouard2010debye, wieth2014dynamical, seyedi2016dynamical, atawa2019molecular, holzl2021dielectric, koperwas2022computational, alvarez2023debye, henot2023orientational, pabst2025salt}. Notwithstanding, this problem is made difficult by the fact that eq.~\ref{eq_epsw_gKt} applies specifically to the modeling of the experimental problem sketched in Fig.~\ref{fig:fig0}a: a microscopic volume, whose dynamics is characterized by $g_\mathrm{K}(t)$, embedded in an infinite continuous medium of permittivity $\epsilon(\omega)$~ \cite{neumann1986computer}. In contrast, MD simulations necessarily involve finite systems. Using periodic boundary conditions requires dealing with the long-range nature of electrostatic interactions~\cite{neumann1983dipole, cox2020dielectric}, which in turn affects the electrostatic boundary conditions (BCs) of the problem. For instance, in the so-called reaction field method~\cite{neumann1983dipole}, charges further apart than a cut-off are replaced by a continuous medium of permittivity $\epsilon_{RF}$. Another popular approach, illustrated by Fig.~\ref{fig:fig0}b,  evaluates electrostatic interactions with the Ewald summation method and corresponds to conducting (sometimes called \textit{tin foil}) BCs~\cite{de1980simulation, caillol1992asymptotic, boresch1997presumed, kolafa2014static, zhang2016JPCL}, which is equivalent to $\epsilon_{RF} \rightarrow \infty$. As $\epsilon(\omega)$ is by definition independent of the electrostatic BCs, dipolar correlations as well as their dynamics need to adapt. Likewise, choosing $\epsilon_\mathrm{RF} \approx \epsilon$ directly reproduces the static correlation $g_\mathrm{K}$~\cite{neumann1983dipole} but leads to a faster dynamics~\cite{neumann1984consistent} than the infinite system. Conversely, with $\epsilon_{RF}\rightarrow \infty$, dipolar correlations are not directly reproduced. This justifies the introduction of a finite-system Kirkwood factor~\cite{rahman1971molecular, neumann1980influence} $G_\mathrm{K} = \langle M^2\rangle /(N\mu^2)$, where $\vec{M}$ is the total dipole moment of the system containing $N$ molecules. This definition is the same as its infinite-system counterpart $g_\mathrm{K}$ (see eq.~\ref{eq_def_gKt}), but corresponds to different BCs. While $g_\mathrm{K}$ is unique to the experimental case, it exists as many values of $G_\mathrm{K}$ as choices of $\epsilon_\mathrm{RF}$. In the following, we will consider only the case $\epsilon_\mathrm{RF}\rightarrow\infty$. In the non-polarizable case ($\epsilon_\infty = 1$), to which we will restrict ourselves from now on, both quantities are related by~\cite{neumann1983dipole}:
\begin{align}
    G_\mathrm{K} = \frac{3\epsilon}{2\epsilon+1}g_\mathrm{K}
    \label{eq_GKgK}
\end{align}
On the other hand, the link between $\epsilon(\omega)$ and $G_\mathrm{K}(t)$ is much simpler for $\epsilon_\mathrm{RF}\rightarrow\infty$ than for the infinite system~\cite{neumann1984consistent}, \textit{viz.}
\begin{align}
    \epsilon(\omega)-1 = \lambda(G_\mathrm{K}-i\omega\mathcal{L}_{i\omega}[G_\mathrm{K}(t)])
    \label{eq_epsw_GKt}
\end{align}
These considerations ensure that the comparison between experiments and simulations regarding $\epsilon(\omega)$ is valid. However, we have to keep in mind that the microscopic dynamics and the amplitude of the cross-correlations depend \textit{a priori} on the choice of electrostatic BCs. Besides, the question of how to disentangle self and cross correlations from $\epsilon(\omega)$ remains. In the recent literature, different approaches have been used to extract a self-part of the dielectric spectra by extrapolating to a single dipole the method used to determine the total spectrum. This has been done either with the standard method considering the total dipole moment of the simulation box~\cite{carlson2020exploring, holzl2021dielectric, pabst2025salt}, or by considering only the restriction of cross correlations to short range~\cite{alvarez2023debye, henot2023orientational, henot2025emergence}. The fact that these methods give a close but not identical result raises the question of the origin of these differences.

In this article, we investigate how dielectric spectra can be practically obtained in MD simulations by considering the dipole moment of a small virtual cavity within a comparatively large simulation box. For three systems of various dielectric permittivities: water, 1-propanol, and diethylether, we systematically compared this approach with the standard one relying on the total dipole moment of the simulation box. We first focus on the static case, for which the short-range cross-correlations can be separated from those induced by the electrostatic BCs used in the simulation. We then test the assumption that the dynamics of the cavity is comparable to that of a cavity embedded in an infinite medium of the same system. We show that this procedure allows for the obtaining of a dielectric spectrum fully compatible with the standard method~\cite{neumann1983dipole}. We discuss the advantages of the cavity method regarding the required simulation time. Finally, we compare different ways of extracting the self-part of a dielectric spectrum, and we discuss how the use of a cavity can help get rid of the effect of electrostatic BCs on the dynamics.

\section{Methods}
MD simulations were performed using OpenMM~\cite{openmm_2017} for water (H$_2$O), 1-propanol (PrOH), and diethylether (Et$_{2}$O) in a cubic box of size $a$ and with PBCs. All-atoms, non-polarizable force-fields, with bonds involving hydrogen constrained, were used as indicated in Table~\ref{table1} alongside with temperature $T$ and system size $N$. Electrostatic interactions were treated using the particle mesh Ewald algorithm~\cite{essmann1995smooth}.  
Simulations were performed with a time step of 2~fs. Each system was equilibrated in the NPT ensemble with a Nosé-Hoover thermostat and a Monte Carlo Barostat at atmospheric pressure. Production runs corresponding to up to 1000 orientational relaxation times were performed in the NVT ensemble. Spectra obtained following the standard procedure of eq.~\ref{eq_epsw_GKt} were computed using the method developed by Bone~\textit{et al.}~\cite{bone2024new}. All other spectra were computed from averaged correlation functions using the fftlog algorithm~\cite{Hamilton_2000}. 

\begin{table}
\caption{\label{table1} Parameters of the simulations for the three systems studied.}
\begin{ruledtabular}
\begin{tabular}{cccccc}
System & Force Field & $N$  & $T$ (K) & $\mu$ (D)  & $\lambda$\\
\hline
H$_2$O &   TIP4P-EW & 13824 & 300 & 2.32 & 18\\
 &  \cite{horn2004development}& / 43950  & & & \\
PrOH & OPLS-AA/3SSPP & 2400 & 340 & 2.88 & 5.4\\
 & \cite{jorgensen1996development, alva2022improving}&  / 9600  & & & \\
Et$_{2}$O  & OPLS-AA~\cite{jorgensen1996development} & 8000 & 280 & 1.48 & 1.4\\
\end{tabular}
\end{ruledtabular}
\end{table}

\section{Static dielectric permittivity}

\begin{figure*}
\includegraphics[width=\linewidth]{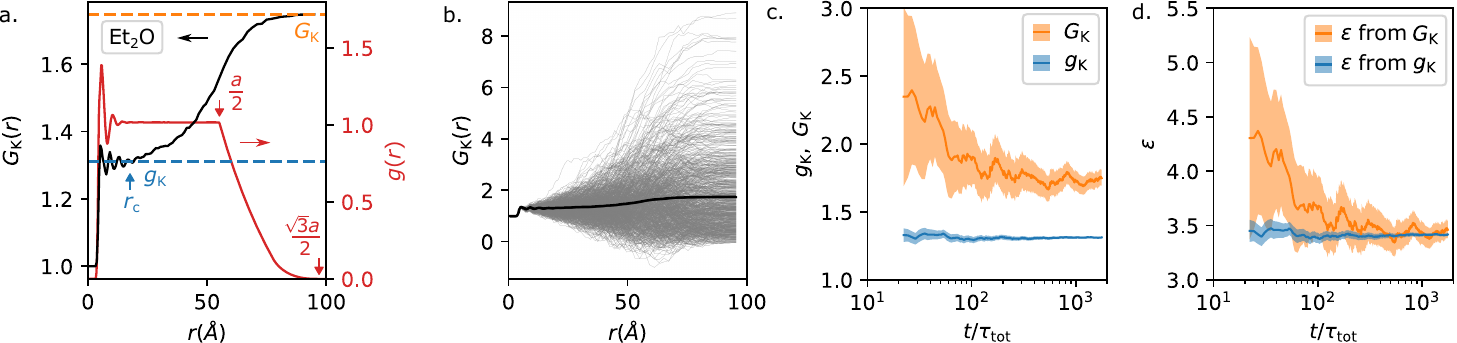}
\caption{\label{fig:fig1}Static cross-correlation for Et$_{2}$O at 280~K. (a) $r$-dependent Kirkwood correlation factor (black, left scale) and pair-correlation factor of the center of charges (red, right scale). The infinite $g_\mathrm{K}$ and finite $G_\mathrm{K}$ correlation factors are shown in blue and orange. (b) Same as (a) for individual snapshots of the simulation. (c) Convergence of $g_\mathrm{K}$ and $G_\mathrm{K}$ as a function of the simulation duration. (d) Same as (c) for the static permittivity obtained from $G_\mathrm{K}$ (eq.~\ref{eq_epsw_GKt}) or from $g_\mathrm{K}$ (eq.~\ref{eq_epsw_gKt}).}
\end{figure*}

In the simulation, having access to the position $\vec{r}_i$ and values $\vec{\mu}_i$ of the dipole moment of each molecule at each time step, we can compute a $r$-dependent Kirkwood correlation factor by restricting the sum of the cross-correlation to a sphere of radius $r$ around a reference dipole $i$ on which we can then average:
\begin{align}
       G_\mathrm{K}(r) = \frac{1}{\mu^2}\left\langle\vec{\mu}_i\cdot\sum_{(r_{ij}<r)}\vec{\mu}_j\right\rangle
       \label{eq_defGK}
\end{align}
This quantity is shown for Et$_2$O in Fig.~\ref{fig:fig1}a as a black curve. It displays a clear increase at the first neighbor distance (see the pair-correlation function of the center of charges in red), followed by a few oscillations of decreasing amplitude. A second significant increase follows with a maximum slope at the half box size $r=a/2$, before reaching its final value $G_\mathrm{K}$ at the half long diagonal of the simulation box $r=\sqrt{3}a/2$.

The static dielectric permittivity $\epsilon$ can be obtained from $\Delta\epsilon = \lambda G_\mathrm{K}$. While this quantity is easy to compute from $\langle M^2\rangle$, it converges notoriously slowly~\cite{fennell2012simple, elton2014polar, zhang2016PRE}. This is illustrated in Fig.~\ref{fig:fig1}b that shows $G_\mathrm{K}(r)$ from individual snapshots of the simulation (\textit{i.e.} averaged over all dipoles $i$ of the simulation box but before averaging over the duration of the simulation). We see that the final value $G_\mathrm{K}$ varies by a few times its mean value. The convergence of its mean and the deduced dielectric permittivity are shown in Figs.~\ref{fig:fig1}c and d as a function of the simulation time. Several hundred relaxation times are needed to accurately determine $\epsilon$. For water at 300 K, this typically requires a 10 ns long simulation~\cite{fennell2012simple, elton2014polar}.

\begin{figure}[h]
\includegraphics[width=\linewidth]{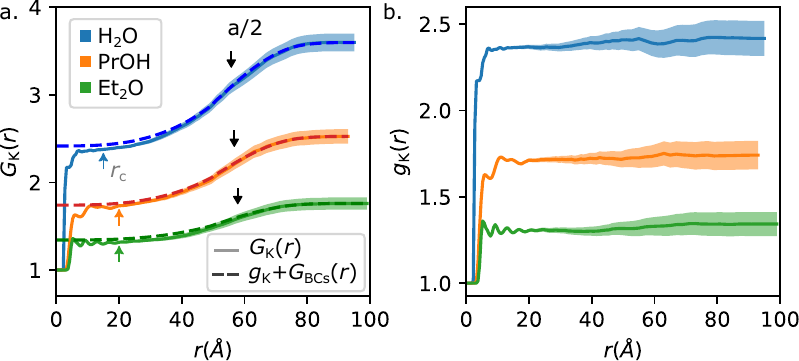}
\caption{\label{fig:fig2}(a) $r$-dependent Kirkwood correlation factor $G_\mathrm{K}(r)$ for three molecular liquids. For each, the contribution $G_\mathrm{BCs}(r)$ given by eq~\ref{eq_G_BCs} is shown with dashed lines. (b) Correlation factor corrected from finite-size effect $g_\mathrm{K}(r) = G_\mathrm{K}(r)  - G_\mathrm{BCs}(r)$.}
\end{figure}

\begin{table}
\caption{\label{table2} Results of the simulation regarding static dielectric quantities.}
\begin{ruledtabular}
\begin{tabular}{ccccc}
System & $G_\mathrm{K}$ & $r_\mathrm{c}$ (\AA) & $g_\mathrm{K}$ & $\Delta\epsilon = \epsilon-1$ \\
\hline
H$_2$O at 300~K    & $3.6 \pm 0.1$ & 15 & $2.36 \pm 0.02$  &  64.8 \\
PrOH at 340~K      & $2.5 \pm 0.1$ & 20 & $1.71 \pm 0.02$  &  13.7 \\
Et$_{2}$O at 280~K & $1.76 \pm 0.1 $& 20 & $1.30 \pm 0.02$  &  2.5 \\
\end{tabular}
\end{ruledtabular}
\end{table}

In the configuration of the simulation, a substantial fraction of the total dipolar correlations (up to 1/3 for a highly polar system) are a consequence of the $\epsilon_\mathrm{RF}\rightarrow\infty$ BCs and are not present in the infinite system. The $r$-dependent finite-system Kirkwood correlation factor can be decomposed into two parts:
\begin{align}
       G_\mathrm{K}(r) = g_\mathrm{K}(r) + G_\mathrm{BCs}(r)
       \label{eq_GKgKGBC}
\end{align}
where $g_\mathrm{K}(r)$ would be obtained from an infinite system, $G_\mathrm{BCs}(r)$ encompasses the consequences of using different electrostatic BCs and was expressed analytically by Caillol~\cite{caillol1992asymptotic}. For $\epsilon_\mathrm{RF}\rightarrow\infty$, it is~\cite{caillol1992asymptotic, zhang2016JPCL}:
\begin{align}
    G_\mathrm{BCs}(r) = \frac{(\epsilon-1)^2}{3\lambda\epsilon}\frac{V(r)}{a^3} 
    \label{eq_G_BCs}
\end{align}
where $V(r)/a^3$ is the volume fraction of the intersection between a sphere of radius $r$ and the cubic simulation box. Knowing this contribution allows one to subtract it from $G_\mathrm{K}(r)$ and recover $g_\mathrm{K}(r)$. This is shown in Fig.~\ref{fig:fig2} for water, propanol, and diethylether. Eq.~\ref{eq_G_BCs} was evaluated without any adjustable parameter by computing $\epsilon$ from $G_\mathrm{K}$. It reproduces perfectly the second increase of $G_\mathrm{K}(r)$ and clearly illustrates that this effect is due to PBCs. Indeed, the flattening of $G_\mathrm{K}(r)$ for $r>a/2$ is related to $V(r)$ being the volume of a cube with side length $a$ truncated by a sphere of radius $r$~\cite{caillol1992asymptotic}. For all three systems, $g_\mathrm{K}(r)$ quickly reaches its final value $g_\mathrm{K}$. This observation is consistent with the work on water of Zhang~\textit{et al.}~\cite{zhang2016JPCL, zhang2016PRE} who demonstrated that the effect of BCs could be removed by combining simulations equivalent to $\epsilon_\mathrm{RF} \rightarrow \infty$ and $\epsilon_\mathrm{RF} \rightarrow 0$.

\begin{figure}
\includegraphics[width=\linewidth]{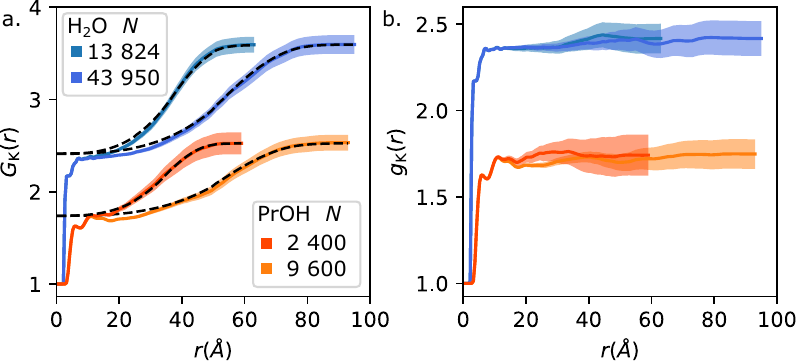}
\caption{\label{fig:fig3}Effect of the number of molecules $N$ in the simulation box for H$_2$O and PrOH on the $r$-dependent Kirkwood correlation factor directly computed from the simulation (a) and corrected from finite-size effects (b).}
\end{figure}
The effect of the simulation box size is shown in Fig.~\ref{fig:fig3}. As predicted by eq.~\ref{eq_G_BCs}, $G_\mathrm{K}$ does not depend on the box size, which can be taken advantage of to accurately determine dielectric spectra with only a few hundred molecules~\cite{holzl2021dielectric}. The effect of electrostatic BCs is not one that can be eliminated by increasing the size of the system. But, if the simulation box is large enough, the two contributions can be spatially separated~\cite{elton2014polar, henot2023orientational}: at short range $G_\mathrm{K}(r)$ is dominated by the evolution of $g_\mathrm{K}(r)$ while the effect of BCs takes over only when $V(r)/a^3$ in eq.~\ref{eq_G_BCs} start to become significant. It is thus possible to carefully choose for each system a cut-off radius $r_\mathrm{c}$ (see Table~\ref{table2}) so that $G_\mathrm{K}(r<r_\mathrm{c}) \approx g_\mathrm{K}(r)$. Indeed, by ensuring $(r_\mathrm{c}/a)^3\ll 1$, we minimize the influence of BCs while conserving the small-scale increase of cross-correlations relevant for an experimental infinite system. This defines a cavity sketched in Fig.~\ref{fig:fig0}c, within a comparatively large but still microscopic medium of permittivity $\epsilon$, itself embedded in a conductive shell. With eq.~\ref{eq_epsw_gKt} and assuming $g_\mathrm{K}\approx G_\mathrm{K}(r_\mathrm{c})$, the dielectric permittivity can be estimated from:
\begin{align}
    \frac{(\epsilon-1)(2\epsilon+1)}{\epsilon} = 3 \lambda G_\mathrm{K}(r_\mathrm{c})
    \label{eq_eps_from_gk}
\end{align}
The values of $g_\mathrm{K}$ and $\epsilon$ obtained this way are given in Table~\ref{table2}, and their convergence is shown in Figs.~\ref{fig:fig1}c and d in blue. We see that using a cavity to get $\epsilon$ gives a result perfectly compatible with the standard method relying on $G_\mathrm{K}$. It should be noted that the cavity method has some drawbacks: it requires simulating a large number of molecules so that the cavity remains small compared to the total system, which may be impractical for polarizable systems. Moreover, obtaining $G_\mathrm{K}(r_\mathbf{c})$ from MD trajectories requires much more computational power than $G_\mathrm{K}$. On the other hand, the cavity method appears to be much less sensitive to noise than the standard one. Indeed, the slow convergence of $G_\mathrm{K}$ seems to arise from $G_\mathrm{BCs}(r)$ rather than $g_\mathrm{K}(r)$. Consequently, accurate values of $\epsilon$ can be obtained from only several tens of total relaxation times instead of several hundred. This can be of interest when studying the dielectric response of liquids at lower temperatures, where relaxation times become much longer. Still, it has to be balanced with the larger simulation box required. For instance, in the present work, simulating a system containing 800 propanol molecules rather than 9600 is 5.5 times faster, which is not far from compensating for the gain in required simulation time. In practice, the benefits of the virtual cavity method become truly significant when a large simulation box is needed to avoid finite-size effects or if one wants to break down the cross-correlations as a function of distance to study how they build up. We previously took advantage of this~\cite{henot2023orientational} to reach low enough temperatures to separate the main relaxation peak from the THz range dynamics by up to six orders of magnitude, enabling precise characterization of the orientational dynamics.

\section{Dielectric spectra}
\begin{figure}
\includegraphics[width=\linewidth]{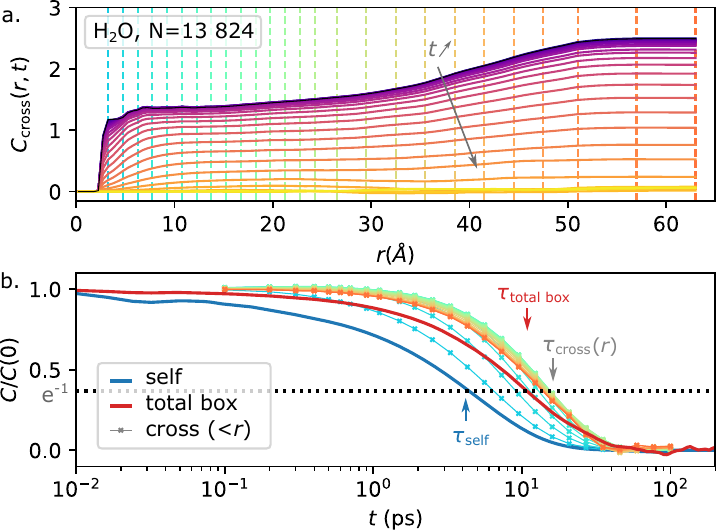}
\caption{\label{fig:fig4} (a) $r$-dependent cross-correlation for water at different log-spaced times (increasing from purple to yellow). (b) Normalized correlation function for the reorientation of individual dipoles $C_\mathrm{self}(t)$ (blue) and of the total simulation box dipole moment $G_\mathrm{K}(r=\sqrt{3}a/2,t)$ (red). The cross part of the correlations within a sphere of radius $r$ is shown for values of $r$ increasing from blue to orange, shown as dashed lines in (a).}
\end{figure}

As in the static case, the time-dependent generalization of the Kirkwood correlation factor (see eq.~\ref{eq_def_gKt}) can also be restricted to a sphere of radius $r$ and decomposed into self $C_\mathrm{self}(t)$ and cross $C_\mathrm{cross}(r, t)$ correlation functions:
\begin{align}
    G_\mathrm{K}(r, t) &= C_\mathrm{self}(t) + C_\mathrm{cross}(r, t)\\
    C_\mathrm{self}(t) &= \frac{1}{\mu^2}\left\langle \vec{\mu}_i(0)\cdot\vec{\mu}_i(t)\right\rangle \\
    C_\mathrm{cross}(r, t) &= \frac{1}{\mu^2}\left\langle \vec{\mu}_i(0)\cdot \sum_{0<r_{ij}<r}\vec{\mu}_j(t)\right\rangle
    \label{eq_defGKrt}
\end{align}

\begin{figure}
\includegraphics[width=\linewidth]{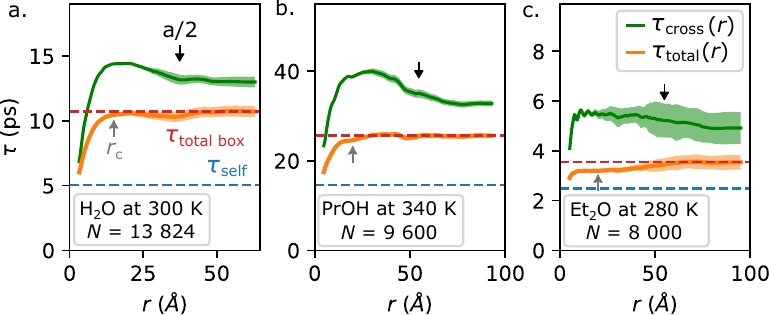}
\caption{\label{fig:fig5} For water (a), propanol (b), and diethylether (c), relaxation time of the total dipole within a sphere of radius $r$ (red markers) and of the cross correlations only (green markers). Horizontal dashed lines correspond to the correlation time of the self (blue) and of the total simulation box (red).}
\end{figure}

It should be noted that the radius $r$ is not as well defined as in the static case, as both molecules $i$ and $j$ have moved between time $0$ and $t$. One solution~\cite{alva2022improving} is to consider, at each time $t$, all molecules $j$ within the sphere of radius $r$ around molecule $i$ ($r_{ij} = |\vec{r}_i(t) - \vec{r}_j(t)|$). In this case, molecules are allowed to enter or leave the virtual cavity, and their number can fluctuate. Another way is to consider only the molecules that were in the sphere at $t=0$ ($r_{ij} = |\vec{r}_i(0) - \vec{r}_j(0)|$), although some of them may be further apart than $r$ at time $t$. In both cases, the correlation function is then averaged over all molecules $i$ of the box and for initial times distributed over the total duration of the simulation. We have checked, for all three systems, that evaluating the distance at time $t$ rather than at $t=0$ leads only to negligible changes in $C_\mathrm{cross}(r,t)$ (see suppl. mat.), and we have chosen, in the following, to compute $r_{ij}$ at $t=0$. The cross correlation functions corresponding to different times are shown for water as a function of $r$ in Fig.~\ref{fig:fig4}a. The already well visible time decorrelation is shown normalized for different radius $r$ in Fig.~\ref{fig:fig4}b, together with the self and the total box correlation functions. From each of these, we define a correlation time $\tau = \int_0^\infty C(t)/C(0)\mathrm{d}t$, shown for all three systems in Figs.~\ref{fig:fig5}a-c. We see that the dynamics of the cross correlations (green curve) is always slower than that of the self (blue dashed line) and depends on the radius $r$. It increases quickly at small $r$, reaching a maximum before decreasing slightly.
The correlation time of the total dipole within a sphere of radius $r$ is given by the weighted average of the self and cross relaxation times.
\begin{align}
    \tau_\mathrm{total}(r) = \frac{1}{G_\mathrm{K}(r)}\tau_\mathrm{self} + \left(1-\frac{1}{G_\mathrm{K}(r)}\right)\tau_\mathrm{cross}(r)
\end{align}
This quantity is shown as a orange curve in Figs.~\ref{fig:fig5}a-c. For water and propanol, in contrast with $\tau_\mathrm{cross}(r)$, it displays only a short-scale increase and quickly reaches its final plateau at $\tau_\mathrm{total~box}$. For diethylether, a first plateau is reached, followed by a small increase at the half box size. In all cases, we can check that at $r=r_\mathrm{c}$ the short-scale increase already took place, which means that the cavity is large enough to encompass the whole dynamics of cross-correlations we are interested in.

\begin{figure*}
\includegraphics[width=\linewidth]{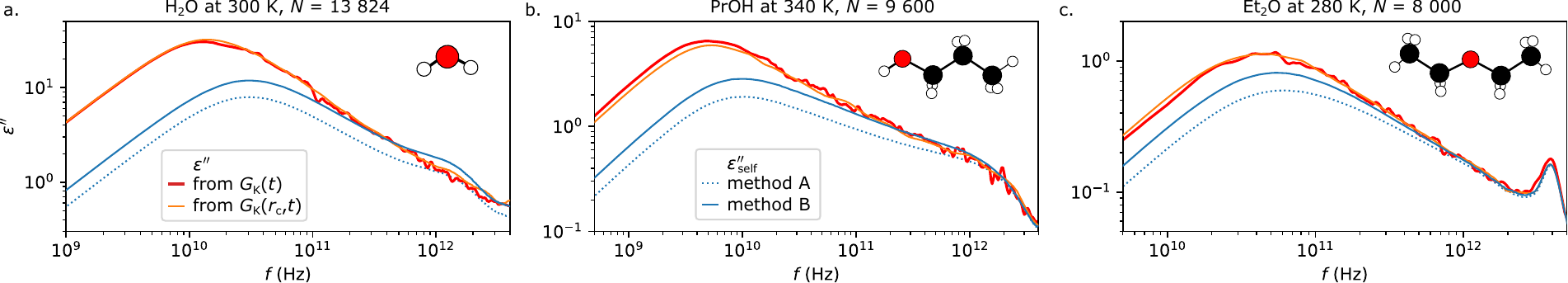}
\caption{\label{fig:fig6} Imaginary part of the dielectric permittivity for water (a), propanol (b) and diethylether (c) computed either from the total box dipole moment from eq.~\ref{eq_epsw_GKt} (red curve), or by considering a cavity of radius $r_\mathrm{c}$ from eq.~\ref{eq_epsw_GKt_rc} (orange curve). The self-part of the dielectric spectra is shown, either obtained from method A (eq.~\ref{eq_epsw_self_A}, dashed blue curve) or from method B (eq.~\ref{eq_epsw_self_B}, solid blue curve).}
\end{figure*}

In the previous section, we showed that $\epsilon$ could be obtained from $G_\mathrm{K}(r_\mathrm{c})$, and we would like to estimate $\epsilon(\omega)$ in the same spirit. This requires to make the assumption that $g_\mathrm{K}(t)$ can be approximated by $G_\mathrm{K}(r_\mathrm{c},t)$, \textit{i.e.} that the dipole moment of a cavity of radius $r_\mathrm{c}$ (see Fig.~\ref{fig:fig0}c) in the simulation behaves as if the cavity were surrounded by an infinite medium of permittivity $\epsilon(\omega)$. With that, $\epsilon(\omega)$ is related to $G_\mathrm{K}(r_\mathrm{c},t)$ by:
\begin{align}
    \frac{(\epsilon(\omega)-1)(2\epsilon(\omega)+1)}{\epsilon(\omega)} = 3\lambda(G_\mathrm{K}(r_\mathrm{c},0)-i\omega\mathcal{L}_{i\omega}[G_\mathrm{K}(r_\mathrm{c}, t)])
    \label{eq_epsw_GKt_rc}
\end{align}
This relation can be inverted to obtain the permittivity $\epsilon(\omega) = \Sigma(\omega) + \sqrt{1+[\Sigma(\omega)]^2}$, with $\Sigma(\omega) = 1/4 + 3\lambda/4(G_\mathrm{K}(r_\mathrm{c})-i\omega\mathcal{L}_{i\omega}[G_\mathrm{K}(r_\mathrm{c}, t)])$. The resulting spectra $\epsilon^{\prime\prime}(\omega)= -\mathrm{Im}[\epsilon(\omega)]$ are shown with orange curves in Figs.~\ref{fig:fig6}a-c for all three systems. They can be compared to the result obtained from the standard method of eq.~\ref{eq_epsw_GKt}, that is to say directly from the Fourier transform~\cite{bone2024new} of the simulation box dipole correlation function (red curves). For each system, both methods give the same result. This validates the assumption that everything happens as if the cavity of radius $r_\mathrm{c}$ was embedded in an infinite medium of permittivity $\epsilon(\omega)$. And indeed, in the present simulations, the fraction of molecules within the cavity with respect to the total simulation box is only 3.4~\% for water and 2.5~\% for propanol and diethylether. This is small enough so that $G_\mathrm{BCs}(r)$ remains negligible. As in the static case, the virtual cavity method allows for obtaining a dielectric spectrum from a significantly shorter simulation run compared to using the total dipole moment of the simulation box. 

To reduce the computational cost of the simulation, we have neglected the effect of electronic polarizability, like many MD studies of this kind~\cite{saiz2000dielectric, atawa2019molecular, alvarez2023debye}. It is worth mentioning the consequences of this simplification. For water, polarizable and non-polarizable models have been directly compared in the literature~\cite{elton2014polar, carlson2020exploring, bone2024new}, showing that taking into account polarizability is required to reproduce the experimentally observed features of dielectric spectra above $10^{12}$~Hz. In contrast, the main relaxation peak was shown to be not much affected by polarizability effects. When focusing on the structural relaxation, this makes non-polarizable models a reasonable trade-off by enabling larger systems to be simulated over longer periods.

The permittivity $\epsilon^{\prime\prime}(\omega)$ can be obtained in a more straightforward way by assuming that $\epsilon(\omega)\gg 1/2$, leading to: 
\begin{align}
    \epsilon(\omega)-1 \approx \frac{3}{2}\lambda(G_\mathrm{K}(r_\mathrm{c},0)-i\omega\mathcal{L}_{i\omega}[G_\mathrm{K}(r_\mathrm{c}, t)])
    \label{eq_epsw_GKt_rc2}
\end{align}
The result is shown as dashed orange curves in Figs.~\ref{fig:fig7}d-f and appears reasonably close to the result of eq.~\ref{eq_epsw_GKt_rc} when focusing on the main relaxation peak for water and propanol.  This is the method we used in previous work on glycerol~\cite{henot2023orientational, henot2025emergence} and up to a prefactor it is equivalent to the approach chosen by Alvarez~\textit{et al.}~\cite{alvarez2023debye} on water who consider a cavity of radius $r_\mathrm{c}=40$~\AA~in a giant simulation box ($a=333$~\AA). For diethylether, however, the approximation is not valid due to the dielectric permittivity being close to unity, and eq.~\ref{eq_epsw_GKt_rc} has to be used instead. This illustrates how the electrostatic BCs can affect the microscopic dynamics: the dipole moment of a small microscopic cavity decorrelates 10~\% faster and less exponentially that the box total dipole moment which gives the correct dielectric spectra (see $C(t)/C(0)$ in Figs.~\ref{fig:fig7}a-c).

\section{Self part of a dielectric spectrum}

\begin{figure*}
\includegraphics[width=\linewidth]{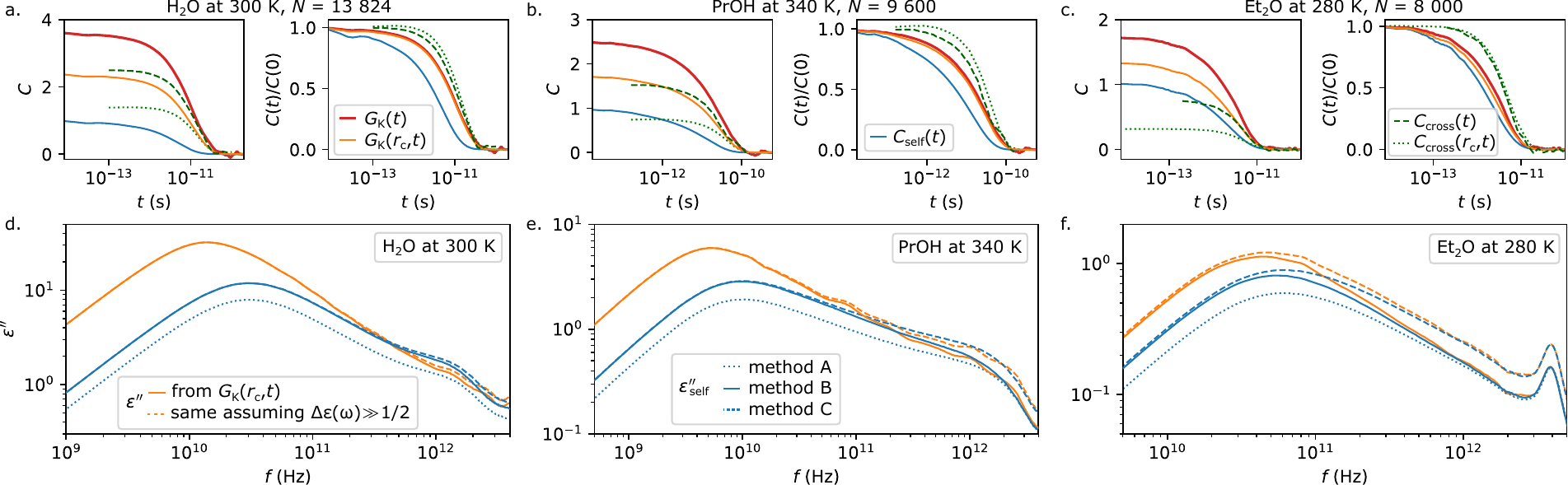}
\caption{\label{fig:fig7}For water (a,d), propanol (b,e), and diethylether (c,f), dipole correlation functions (a-c) are shown for the total simulation box (red), for the cavity or radius $r_\mathrm{c}$ (orange) and, for the self part (blue). The cross-part is shown in green for the cavity (dotted line) and the whole box (dashed line). The dielectric spectra (d-f) obtained from eq.~\ref{eq_epsw_GKt_rc} and \ref{eq_epsw_GKt_rc2} are shown with solid and dashed orange lines, respectively. The self dielectric spectra obtained from method A (eq.~\ref{eq_epsw_self_A}, dotted line), B (eq.~\ref{eq_epsw_self_B}, solid line), and C (eq.~\ref{eq_epsw_self_C}, dashed line) are shown in blue.}
\end{figure*}

The imaginary part $\epsilon^{\prime\prime}(\omega)$ of the dielectric response is a convenient practical representation: it is directly accessible experimentally, has a clear physical meaning, and makes it easy to separate relaxation processes. When attributing different features of a spectrum to specific microscopic mechanisms, it can be useful to decompose it into a sum of (a priori non-independent) processes, each having its spectral shape: for example, a Debye process, an $\alpha$ peak, a secondary process, and so on. By combining different experimental techniques, or with MD simulations, it can be possible to disentangle these processes. As detailed above, it is in particular the case for the self and cross dipole dynamics as visible in Figs.~\ref{fig:fig6}a-c through the decomposition $G_\mathrm{K}(t) = C_\mathrm{self}(t) + C_\mathrm{cross}(t)$. Similarly, it is tempting to decompose a dielectric spectrum into self and cross contributions: $\epsilon^{\prime\prime}(\omega) = \epsilon^{\prime\prime}_\mathrm{self}(\omega) + \epsilon^{\prime\prime}_\mathrm{cross}(\omega)$. However, this decomposition raises a problem rooted in the non-linear relationship between the dielectric permittivity and the microscopic dynamics given by eq.~\ref{eq_epsw_gKt}. Nevertheless, when performing the standard analysis of a MD simulation with $\epsilon_\mathrm{RD}\rightarrow\infty$, this relationship happens to be linear (see eq.~\ref{eq_epsw_GKt}). This gives a straightforward method (that we denote A) to obtain the self part of the dielectric spectrum~\cite{carlson2020exploring, holzl2021dielectric, pabst2025salt} :
\begin{align}
\Delta \epsilon_\mathrm{self, A}(\omega) = \lambda(1-i\omega\mathcal{L}_{i\omega}[C_\mathrm{self}(t)])
    \label{eq_epsw_self_A}
\end{align}
Despite its apparent simplicity, this expression raises questions: the static value $\Delta\epsilon_\mathrm{self,~A}(0)/\Delta\epsilon(0) = 1/G_\mathrm{K}$ only makes sense in the configuration of the simulation with $\epsilon_\mathrm{RF}\rightarrow\infty$ BCs. Yet, we would like this decomposition to be also relevant from an experimental point of view. For instance, in the case of a polar liquid having a temperature such that $g_K = 1$, we would get $\Delta\epsilon_\mathrm{self,~A} \neq \Delta\epsilon$ because in this case $G_\mathrm{K} = 3\epsilon/(2\epsilon+1) \neq 1$. Only in the limiting case $\Delta \epsilon\ll 1$, we would recover the expected behavior. 

One could imagine another approach (denoted B) where the self part of the dielectric spectrum is obtained by replacing the correlation function of the cavity by that of the self dipole. This leads to:
\begin{align}
\epsilon_\mathrm{self, B}(\omega) &= \Sigma_\mathrm{self}(\omega) + \sqrt{1+[\Sigma_\mathrm{self}(\omega)]^2},\\\Sigma_\mathrm{self}(\omega) &= \frac{1}{4} + \frac{3}{4}\lambda(1-i\omega\mathcal{L}_{i\omega}[C_\mathrm{self}(t)]) 
    \label{eq_epsw_self_B}
\end{align}
 Physically, this corresponds to the dielectric permittivity of a material of the same $\lambda$ but for which net cross correlations would vanish ($g_\mathrm{K}=1)$ without affecting its self dynamics. The results from methods A and B are shown in Figs.~\ref{fig:fig6}a-c with dashed and solid lines, respectively. They differ by their amplitude (by a factor close to $G_\mathrm{K}/g_\mathrm{K}$ reaching 3/2 for large $\Delta\epsilon$) and by their spectral shape when $\Delta\epsilon$ is close to 1, as it is the case for diethylether. We note that methods A and B become equivalent in the limit $\lambda\ll 1$ (\textit{i.e} $\Delta \epsilon\ll 1$), that is to say, for a dilute gas of polar molecules or for an almost nonpolar liquid. The fact that the same correlation function $C_\mathrm{self}(t)$ can lead to different spectral shape is related to the nonlinear relationship between $\epsilon(\omega)$ and the correlation function~\cite{fatuzzo1967theory, bottcher1978}. Yet, taking into account this distortion for the self spectrum does not make much sense in the context of the experimental efforts comparing the orientational dynamics characterized by dielectric spectroscopy and by depolarized dynamic light scattering. Indeed, this second technique, much less sensitive to cross-correlations~\cite{ bohmer2024spectral}, can give access to the self dynamics directly through a time correlation function~\cite{gabriel2017debye}. Also, non-linearity leads to the main limitation of method B: it is a non-additive decomposition of $\epsilon(\omega)$. For diethylether, this procedure significantly affects the shape of $\epsilon_\mathrm{self, B}(\omega)$ and removes all meaning from $\epsilon(\omega)-\epsilon_\mathrm{self,B}(\omega)$.
 
For these reasons, it is interesting to consider the limit $\Delta \epsilon(\omega) \gg 1/2$, for which linearity, and thus additivity, are recovered. This is denoted as method C:
 \begin{align}
\Delta \epsilon_\mathrm{self, C}(\omega) = \frac{3}{2}\lambda(1-i\omega\mathcal{L}_{i\omega}[C_\mathrm{self}(t)])
    \label{eq_epsw_self_C}
\end{align}
Here the self spectra has the same spectral shape as in method~A but, because the cross correlations beyond $r_\mathrm{c}$ are not taken into account, its static value is 1.5 times larger: $\Delta\epsilon_\mathrm{self,~C}(0)/\Delta\epsilon(0) = 1/g_\mathrm{K}$. It is shown in Fig.~\ref{fig:fig7} as a dashed blue line. For water, we see that it approximates well the result of method B. For propanol, it is acceptable close to the main relaxation peak, and for diethylether it is, as expected, not suited at all. Here the difference $\epsilon_\mathrm{cross}(\omega) = \epsilon(\omega)-\epsilon_\mathrm{self,C}(\omega)$ directly characterizes the dynamics of the short range cross-correlations. In this framework, it is possible to further decompose these cross-correlations as a function of $r$, which can help in the understanding of their physical origin~\cite{alva2022improving, henot2023orientational}. For water $\epsilon^{\prime\prime}_\mathrm{self, C}(\omega)$ appears to cross $\epsilon^{\prime\prime}(\omega)$ at high frequency (see Fig.~\ref{fig:fig7}). This may seems surprising but can be attributed to the fact that the correlation function $C_\mathrm{cross}(r_\mathrm{c},t)$ is not a strictly decreasing function (see Fig.~\ref{fig:fig7}a) so that the $\epsilon^{\prime\prime}_\mathrm{cross}(\omega)$ is not always positive. 

Overall, the difference between the methods detailed above lies in whether or not correlations originating from $r\in[r_\mathrm{c}, \sqrt{3}a/2]$ are included. Including them increases the static correlation from $g_\mathrm{K}$ to $G_\mathrm{K}$ and, as a consequence, modifies the weighting between total and self spectra. In general, short and long-range cross-correlation have distinct dynamics. This effect is maximum when $\Delta\epsilon\approx 1$ (see Fig\ref{fig:fig7}c) but becomes negligible if $\Delta \epsilon \gg 1$ as visible in Figs.~\ref{fig:fig7}a-b, where only the effect on the amplitude remains, or if $\Delta \epsilon\ll 1$. Due to their dependence on the somewhat arbitrary choice of BCs used in the simulation, it seems reasonable not to include those long-range cross-correlations when decomposing a dielectric spectrum into self and cross. Yet, this remains a small effect, and these considerations do not call into question the physical conclusions obtained when using either procedure.

\section{Conclusion}
We have studied in detail how dielectric spectra could be obtained in MD simulations with conductive electrostatic BCs by considering the fluctuation of the dipole in a virtual cavity rather than of the total simulation box. The cavity must be chosen large enough to contain the short-range dipolar correlation present in polar liquids, but small enough compared to the simulation box, so that correlations induced by the BCs remain negligible. We verified that this approach is consistent with the standard method regarding the resulting dielectric spectra for three liquids with dielectric permittivities ranging from low to high. In addition, it provides a direct access to the static and dynamic correlations representative of the experimental infinite case. In practice, it requires simulating a large system and performing a computationally costly analysis of the trajectories, but it has the benefit of requiring significantly less statistics to obtain dielectric spectra than the standard method. This is of practical interest when a very good signal-to-noise ratio is needed, for example, for accessing the non-linear dielectric response, which involves a subtraction between almost identical linear spectra~\cite{henot2025emergence}. 

The cavity method gives an occasion to shed a direct light on the effect of BCs in MD simulation, well known in the literature, but paradoxically concealed behind the simplicity of eq.~\ref{eq_epsw_GKt}. Due to the non-linear relationship between $\epsilon(\omega)$ and the molecular correlation function, there is no general, fully satisfactory way, when using either a virtual cavity or the total box dipole moment, of decomposing a dielectric spectrum into self and cross parts that remains consistent with the infinite system case. Fortunately, this problem can be reasonably overcome when focusing on the main relaxation peak of polar liquids whenever $\Delta\epsilon>10$, which is of practical interest in such non-polarizable classical MD simulations. The cavity approach can then be used to study in details how short-range cross-correlations build-up to better understand their origin. Applying this method, in the case of 1-propanol, to investigate of how they are linked to supramolecular structures formed by hydrogen bonds~\cite{wieth2014dynamical} will be the subject of a future work.

\section*{Supplementary Material}
See the supplementary material (doi.org/10.60893/figshare.jcp.c.8004019.v1) for details on how the choice of method for defining a virtual cavity affects the computation of dynamical cross-correlations.

\begin{acknowledgments}
The author is grateful to P.M. Déjardin, F. Pabst, F. Ladieu and J.P. Gabriel for fruitful discussions and for feedback on the manuscript. Financial support by CEA/DRF/IRAMIS institute and Paris-Saclay university is acknowledged.
This project was provided with computing HPC and storage resources by GENCI at TGCC thanks to the grant 2025-16286 on the supercomputer Joliot Curie's ROME and V100 partition. 
\end{acknowledgments}

%\nocite{*}
\bibliography{biblio}% Produces the bibliography via BibTeX.

\end{document}